\documentclass[prl,showpacs,twocolumn,superscriptaddress]{revtex4-1}
\pdfoutput=1
\usepackage{amsmath,latexsym,natbib,bm,psfrag,color}
\usepackage[pdftex]{graphicx}
\usepackage{epstopdf}

\begin{document}

\title{ Highly-confined spin-polarized two-dimensional electron gas in 
        SrTiO$_{3}$/SrRuO$_{3}$ superlattices }

\author{ Marcos Verissimo-Alves }
\affiliation{ Departamento de Ciencias de la Tierra y
              F\'{\i}sica de la Materia Condensada, Universidad de Cantabria,
              Cantabria Campus Internacional,
              Avenida de los Castros s/n, 39005 Santander, Spain}
\author{ Pablo Garc\'{\i}a-Fern\'andez }
\affiliation{ Departamento de Ciencias de la Tierra y
              F\'{\i}sica de la Materia Condensada, Universidad de Cantabria,
              Cantabria Campus Internacional,
              Avenida de los Castros s/n, 39005 Santander, Spain}
\author{ Daniel I. Bilc }
\affiliation{ Physique Th\'eorique des Mat\'eriaux, 
              Universit\'e de Li\`ege, All\'ee du 6 de Ao\^ut 17 (B5), 
              B-4000 Sart Tilman, Belgium}
\author{ Philippe Ghosez }
\affiliation{ Physique Th\'eorique des Mat\'eriaux, 
              Universit\'e de Li\`ege, All\'ee du 6 de Ao\^ut 17 (B5), 
              B-4000 Sart Tilman, Belgium}
\author{ Javier Junquera }
\affiliation{ Departamento de Ciencias de la Tierra y
              F\'{\i}sica de la Materia Condensada, Universidad de Cantabria,
              Cantabria Campus Internacional,
              Avenida de los Castros s/n, 39005 Santander, Spain}

\date{\today}

\begin{abstract}
 We report first principles characterization of the structural and 
 electronic properties of (SrTiO$_{3}$)$_{5}$/(SrRuO$_{3}$)$_{1}$ superlattices.
 We show that the system exhibits a spin-polarized two-dimensional electron gas,
 extremely confined to the 4$d$ orbitals of Ru in the SrRuO$_{3}$ layer. 
 Every interface in the superlattice behaves as a  
 minority-spin half-metal ferromagnet, with a magnetic moment of 
 $\mu$ = 2.0 $\mu_{\rm B}$/SrRuO$_{3}$ unit.
 The shape of the electronic density of states, half metallicity 
 and magnetism are 
 explained in terms of a simplified tight-binding model, considering 
 only the $t_{2g}$ orbitals plus 
 (i) the bi-dimensionality of the system, 
 and (ii) strong electron correlations.
\end{abstract}


\maketitle


 The family of ABO$_{3}$ perovskite compounds is really prominent 
 among all the complex oxides.
 The delicate interaction between their electronic, spin, lattice and orbital
 ordering degrees of freedom, whose respective energy couplings
 are all of the same order of magnitude~\cite{Rondinelli-11},
 gives rise to a wide variety of ground states and phenomena~\cite{Zubko-11}.
 These compounds typically exhibit exceptional functional properties
 that, moreover, can be finely tuned by playing with temperature and/or 
 electrical (screening)~\cite{Stengel-09-np}, 
 mechanical (strain)~\cite{Schlom-07},
 and chemical (doping)~\cite{RVWang-09} boundary conditions. 

 At the bulk level, ABO$_{3}$ compounds cover the whole spectrum of conducting 
 properties, ranging from good insulators and semiconductors to metals and even 
 superconductors~\cite{Zubko-11}.
 Moreover, it has recently been shown that their conducting behaviour can also 
 drastically change at  surfaces~\cite{Santander-11} and 
 interfaces~\cite{Ohtomo-02}. The possibility to create 
 highly-confined two-dimensional electron gases (2DEGs) at oxide 
 interfaces has also been demonstrated and generated a huge excitement.

 The most widely studied system is certainly the 2DEG formed at 
 the LaO/TiO$_2$ polar 
 interface between LaAlO$_3$ (LAO) and SrTiO$_3$ (STO)~\cite{Ohtomo-04}, 
 two good band insulators at the bulk level. 
 Although the mechanism at the origin of the 
 transfer of charges remains partly controversial, the 2DEG is 
 formed by the accumulation 
 of electrons in the Ti 3$d$ states of STO over a thickness 
 of a few nanometers near the interface,
 and exhibits interesting exotic properties.  
 In a different context, when looking for oxides with enhanced 
 thermoelectric properties,  
 Ohta {\it et al.}~\cite{Ohta-07-2} highlighted the possibility 
 to create a highly-confined 2DEG in   
 [(STO)$_{m}$/(SrTi$_{0.8}$Nb$_{0.2}$O$_{3}$)$_{n}$]$_{p}$ superlattices.
 Since Nb-doped STO is a well-known conducting 
 electrode material~\cite{Tybell-99},
 such a superlattice can be viewed as the limiting case of 
 ultrathin metal oxide layers embedded
 in an insulating STO matrix. 
 The 2DEG arises from the doping produced by the 
 Nb atoms so, alternatively, the system can also be interpreted as a 
 \emph{partial} doping of a STO matrix at the B site. 
 Very recently, the formation 
 of a 2DEG has been achieved through a related route: the doping of the A site
 of the STO matrix. 
 The Sr atoms of a single SrO layer of STO
 are partially~\cite{Ong-11} or completely~\cite{Jang-11} replaced
 by a rare-earth element, $R$, which provides additional electrons. 
 In this case, the conduction electrons 
 provided by the substitutional layer are transferred to the STO matrix 
 but stay near the $R$O layer due to Coulomb attractions, 
 and possibly also to the large dielectric constant of STO~\cite{Stengel-11.2}.

 In this Letter, we predict from first principles calculations the appearance
 of a highly-confined 2DEG in (STO)$_5$/(SrRuO$_3$)$_1$ superlattices (STO/SRO). 
 This system is to some extent related to those previously considered. 
 SRO is a well-known conductive material so that 
 STO/SRO present similarities with 
 (STO)$_{m}$/(SrTi$_{0.8}$Nb$_{0.2}$O$_{3}$)$_{n}$ 
 superlattices~\cite{Ohta-07-2}. 
 Also, STO/SRO can be seen as a STO matrix  in which a TiO$_2$ layer is 
 periodically replaced by a single RuO$_2$ layer, appearing as a variant 
 of the work of Jang {\it et al.}~\cite{Jang-11}.
 However, contrary to all recent examples of 2DEGs in perovskites previously 
 discussed, that confine the conducting electrons in 
 Ti 3$d$ orbitals of STO, the conducting electrons in STO/SRO are 
 localized \emph{exclusively} in Ru 4$d$ orbitals. 
 We will show that this gives rise to new and unexpected features.
 Contrary to current though regarding SRO 
 \emph{thin films}~\cite{Mahadevan-09,Chang-09,Xia-09},
 and \emph{finite period} superlattice~\cite{Izumi-98},
 our infinite period  superlattices
 exhibits an extremely confined half-metallic ferromagnetic (FM) 
 2DEG, very promising for spintronic devices~\cite{Mandal-10}.
  

 For this study, we perform first principles simulations
 of STO/SRO superlattices using two complementary approaches. 
 Unless otherwise stated, all the calculations presented here
 have been performed within the local spin 
 density approximation (LSDA) to the density functional theory 
 using the {\sc Siesta} code~\cite{Soler-02}.
 An extra Hubbard-U term
 is included to account for the strong electron correlations,
 with a $U_{\rm eff}$ of 4.0 eV applied only to the $d$ orbitals of Ru.
 The robutness of the results presented below
 have been doubled checked using 
 the {\sc Crystal09} code~\cite{crystal-web} 
 within the B1-WC hybrid scheme~\cite{Bilc-08}.
 Hybrid functionals are well known to provide improved 
 description of magnetic and highly-correlated oxides, 
 and B1-WC appeared recently as a successful approach to  
 describe 2DEG at the LAO/STO interface~\cite{Delugas-11}. 
 The excellent agreement between the results obtained independently 
 within both approaches strongly supports our predictions. 
 The procedure to fit the $U_{\rm eff}$ and other technicalities can be found
 in the Discussion 1 and 2 of the Supplemental materials~\cite{Supplementary}.
 Both spin-orderings in the SRO layers, FM and antiferromagnetic (AFM),
 where the up and down spins of the RuO$_{2}$ layer 
 are ordered in a checkerboard arrangement, 
 are simulated.


 \begin{figure}[htbp]
    \begin{center}
       \includegraphics[width=\columnwidth]{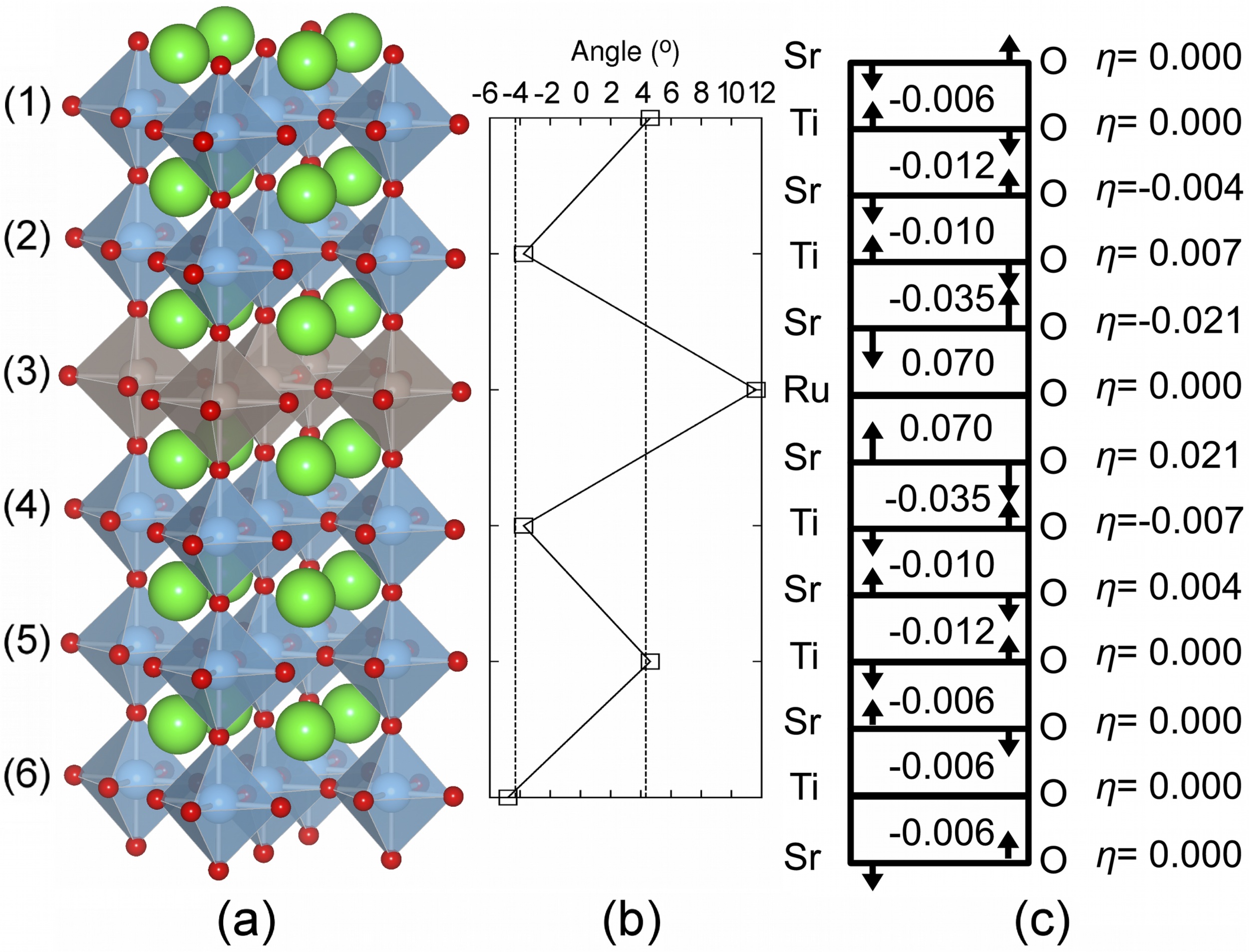}
       \caption{ (color online) 
                 Schematic representation of the relaxed structure of 
                 STO/SRO superlattices with LDA+U. 
                 (a) Unit cell periodically repeated in space. 
                 O atoms are shown in red, Sr in green, Ti in blue and 
                 Ru in grey, with O octahedra colored accordingly. 
                 (b) Rotation angle of the oxygen octahedra, $\theta$, 
                 around the [001] direction ($z$ axis). 
                 Squares represent $\theta$ 
                 for the corresponding octahedra
                 at the same height as in panel (a). 
                 Dotted lines are the theoretical value of $\theta$ for 
                 bulk STO. 
                 (c) Layer-by-layer rumpling, 
                 $\eta_{i} = [z ({\rm M}_{i})- z({\rm O}_{i})]/2$, 
                 where $z({\rm M}_{i})$ is the position of  
                 the cation, and $z({\rm O}_{i})$ is the position of O, in 
                 a given layer $i$. Atoms in layer $i$ move in the direction 
                 indicated by the arrows. Black lines represent the mean 
                 position of each atomic layer. Numbers inside the structure 
                 are the change in the interplanar distance between 
                 consecutive planes with respect to the ideal unrelaxed 
                 structure (half of the bulk value of 
                 STO, $a$ = 3.874 \AA ). 
                 Similar results are obtained within the B1-WC fuctional
                 (see Discussion 3 of the Supplemental 
                 material~\cite{Supplementary}).
                 Units in \AA .
               }
       \label{fig:structure}
    \end{center}
 \end{figure}

 In the most stable phase, that turns out to have FM order, 
 our simulations predict a $P4/mbm$ space group. 
 Oxygen octahedra rotate changing the phase from layer to layer, 
 as shown in Fig.~\ref{fig:structure}(b). 
 This pattern of rotation shares some similitudes with the 
 $a^{0}a^{0}c^{-}$ Glazer rotation of the STO matrix,
 but now the module of the rotation angles around the $z$-axis are not
 the same from one octahedra to the adjacent one.
 The angle of rotation of the RuO$_{6}$ octahedra, 
 $\theta_{\rm SRO}$ = 11.6$^\circ$, 
 is roughly the same as in bulk tetragonal 
 SRO under the same strain condition in the basal plane
 ($\theta^{\rm bulk}_{\rm SRO}$ = 11.4$^\circ$). 
 The rotations of the TiO$_{6}$ octahedra rapidly converge to the
 bulk LDA value of STO ($\theta^{\rm bulk}_{\rm STO}$ = 4.3$^\circ$),
 indicating that the rotations are functions of the local chemical 
 environment~\cite{Wu-11}.
 Taking into account that bulk SRO crystallizes within the GdFeO$_{3}$
 structure ($Pnma$ space group), with $a^{-}b^{+}a^{-}$ Glazer rotation,
 we have explored the possibility of the existence of out-of-plane tiltings in 
 SRO. However, under the epitaxial configuration considered here, we have
 checked that they dissappear. 
 This supports that the single unit cell of SRO in our superlattices
 are stabilized in a pseudocubic perovskite structure~\cite{Eom-92}.
 As shown in Fig.~\ref{fig:structure}(c), 
 both the rumplings, and the changes in interplanar spacing 
 with respect to the ideal (non-relaxed) structure, 
 are very small.
  

 \begin{figure*}
    \begin{center}
       \includegraphics[width=17.0cm]{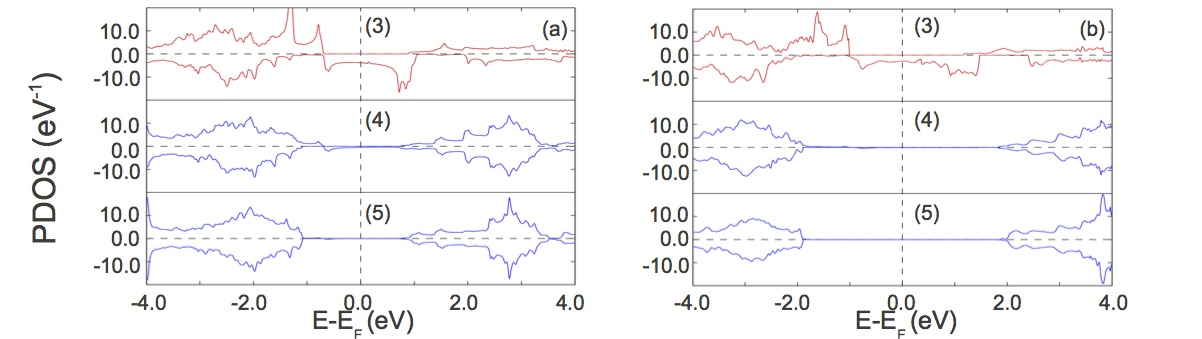}
       \caption{ (color online) 
                 Layer-by-layer PDOS on the atoms at the SrBO$_{3}$ layers 
                 (B = Ti or Ru),
                 within (a) the LSDA+U, and (b) the B1-WC hybrid functional,
                 for the corresponding relaxed STO/SRO
                 superlattice. Each panel represents the PDOS of a unit cell,
                 numbered as in 
                 Fig.~\ref{fig:structure}(a).
                 Corresponding layers in the upper-half of the structure are 
                 equivalent by symmetry.
                 Majority/minority spin are 
                 represented in the upper/lower halves of each panel, 
                 respectively.
               }
       \label{fig:dos}
    \end{center}
 \end{figure*}

 Figure~\ref{fig:dos} shows the projected density of states (PDOS)
 on the atoms at the different SrBO$_{3}$ unit cells (where B is Ti or Ru),
 within the LDA+U [panel (a)], and B1-WC hybrid functional [panel (b)].
 The physical conclusions that
 can be drawn are robust, independently of the theoretical framework used. 
 In particular: 
 (i) remarkably the system presents a 2DEG, strictly confined in the 
 SRO layer. 
 The electronegativity of Ru (2.20), larger than the one for Ti (1.54),
 is the main candidate to
 be the driving force for the electron localization.
 This localization resembles that recently found on the $d$ orbitals of V
 in SrVO$_{3}$/Nb:SrTiO$_{3}$ quantum wells~\cite{Yoshimatsu-11}.
 (ii) The electronic states around the Fermi level are fully 
 spin-polarized, with only the minority-spin electrons involved in the 
 charge transport~\cite{Mandal-10}.
 (iii) The 2DEG exhibits ferromagnetism, with a magnetic moment 
 of $\mu$ = 2.0 $\mu_{\rm B}$/SRO formula unit.
 Integrating the spin-polarized charge density on spheres 
 surrounding the atoms (in {\sc Siesta}) or analyzing the Mulliken
 spin population within B1-WC (in {\sc Crystal}), we see that 
 the magnetic moment is mostly due to the Ru atoms, 
 $\mu_{\rm Ru}$ = 1.4 $\mu_{\rm B}$, 
 within the range of measured magnetic moments
 in single crystals~\cite{Cao-97}, and polycrystalline samples~\cite{Jin-08}. 
 The same value of $\mu_{\rm Ru}$ is obtained in the AFM configuration.
 The remaining magnetic moment in the FM configuration
 comes from the oxygen atoms. 
 These results are robust with respect to pseudopotentials, basis set,
 functional, and  inclusion of spin-orbit coupling (Discussion 4 of 
 the Supplemental material~\cite{Supplementary}). 
 Interestingly, the electronic structure at the interface inherits 
 some of the particularities theoretically predicted in \emph{bulk}
 SRO when the electronic correlations are properly included
 [as in LDA+U~\cite{Jeng-06}, pseudo self-interaction 
 correction~\cite{Rondinelli-08}, or 
 B1-WC hybrid functional (Discussion 5 in the 
 Supplemental materials~\cite{Supplementary}). 
 Between these properties, we might cite the half-metallicity~\cite{Jeng-06} 
 or the saturation of the 
 magnetic moment~\cite{Rondinelli-08}.
 Experimental observation of the half-metallicity remains a challenge,  
 mostly due to the 
 large magnetic fields required to overcome the high magnetic anistropy 
 barriers and fully magnetize the samples.

\begin{figure}[h]
    \begin{center}
       \includegraphics[width=8.0cm]{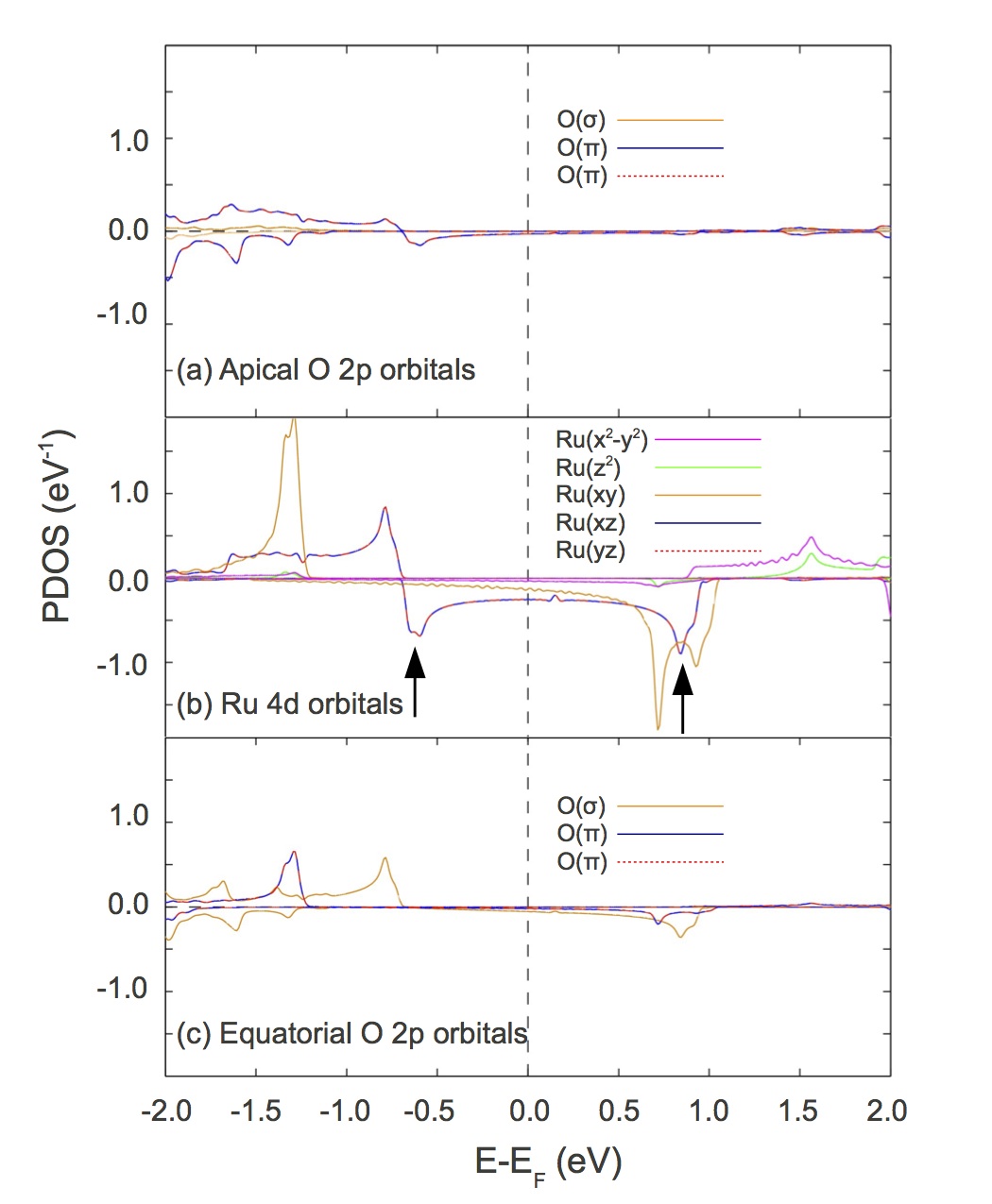}
       \caption{(Color online) PDOS on the atoms at the central SRO 
                layer, projected on the orbitals of 
                (a) an apical O atom, 
                (b) the Ru atom, and
                (c) an equatorial O atom. 
                In the DOS projections corresponding to Oxygen, 
                the 2$p$ orbital directed towards the transition metal ions 
                is denoted by O($\sigma$), while the two quasi-degenerate
                orbitals perpendicular to this bond direction are O($\pi$).
                The arrows in panel (b) mark the position of the peaks for the
                Ru-$d_{xz,yz}$ orbitals (see text).}
       \label{fig:dos-sro}
    \end{center}
 \end{figure}

 Regarding the band shape, we observe in Fig.~\ref{fig:dos-sro} 
 that the bands with Ru-$d_{xy}$ character are significantly different
 from the degenerate Ru-$d_{xz,yz}$ bands. 
 While the PDOS of Ru-$d_{xy}$ resembles that usually found in 
 bulk $t_{2g}$ bands~\cite{Wolfram-06}, 
 the double peaked shape [peaks marked with arrows in Fig.~\ref{fig:dos-sro}(b)]
 for Ru-$d_{xz,yz}$ is similar to that of a 1D metal. 
 It is also significant that the relative position 
 of these bands for the majority and minority spins is different,
 an effect attributable to the strong electronic correlations 
 in the system, enhanced with respect to bulk due to the two-dimensionality 
 of the system (see below).
 Indeed, to study the effect of the Hubbard-U in our calculations 
 we carried out a 
 plain LSDA simulation (without $U$ corrections) and plotted the 
 PDOS (Discussion 2 in the Supplemental materials~\cite{Supplementary}).
 The {\it shape} of the Ru-$d_{xy}$ and Ru-$d_{xz,yz}$ is the same as 
 when the Hubbard-U term is considered.
 However, the position of the bands is shifted to almost the same energy 
 range for both spin channels, destroying the
 half-metallicity.

 \begin{figure}[h]
    \begin{center}
       \includegraphics[width=8.0cm]{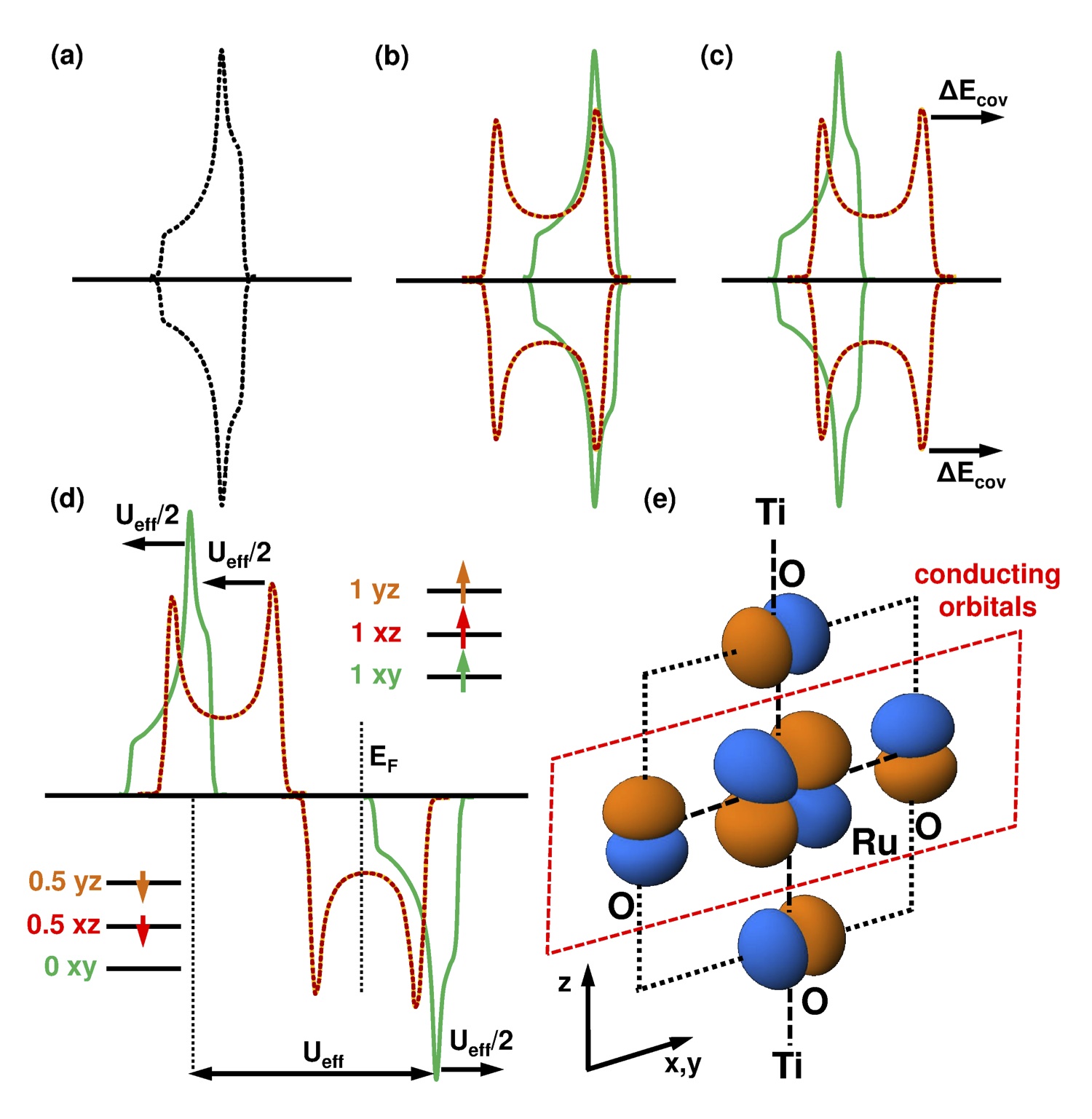}
       \caption{(Color online) Result of the tight-binding model used
                to interpret the DOS of $t_{2g}$ $d$ orbitals.
                Our model shows the transformation 
                from (a) the three-dimensional bulk, 
                to (b) a two-dimensional structure. 
                Refinements of the model to include, first,
                covalent bonding effects in-plane and out-of-plane (c),
                and second, shifts of the bands due to strong 
                electronic correlation via a Hubbard-U term (d),
                are required to reproduce the first principles band structure.
                Black lines in (a) represent the $d_{xy,xz,yz}$ 
                manifold, degenerated in energy. Once the symmetry is 
                broken, the $d_{xy}$ state is represented by a green line,
                and the degenerated $d_{xz,yz}$ orbitals by red lines.
                (e) illustrates 
                the orbitals used in the tight-binding model and
                differences of in-plane 
                and out-of-plane bonding for $d_{xz}$ orbitals 
                (identical sketch can be done for the $d_{yz}$ orbitals). }
       \label{fig:model}
    \end{center}
 \end{figure}

 If the half-metallicity can be attributed to the electronic correlation,
 the shape of the bands can only be explained through the 2D character
 of the electron gas. To check this point, we have made use of a  
 tight-binding model, where only the $t_{2g}$ states are retained in the
 basis set. 
 Qualitatively, the shape of the $t_{2g}$ bands is found to be the same 
 for all perovskites including transition metal ions in the B-position.
 Assuming cubic symmetry, the PDOS associated to these bands for a 
 full three-dimensional solid is shown in Fig.~\ref{fig:model}(a). 
 Using this model we then simulate
 the changes in the PDOS when the system is bidimensional, extended periodically
 through the $xy$ plane. In Fig.~\ref{fig:model}(b) we can see that, 
 while $d_{xy}$-band remains unaltered, the $d_{xz,yz}$ ones now present 
 two-peaks in good agreement with the ones in the full first principles 
 calculations [arrows in Fig.~\ref{fig:dos-sro}(b)]. 
 In particular, the PDOS of bidimensional $d_{xz,yz}$ levels 
 closely resembles that of textbook one-dimensional electron 
 gas~\cite{Wolfram-06} due to the 
 negligible interactions of $d_{xz,yz}$ wavefunctions in the $xy$ plane 
 with other than first neighbors. However, we can observe that the 
 LDA+U bands are shifted with respect to the ones obtained in our 
 tight-binding model. These shifts are due to (i) the difference in bonding 
 of the orbitals with in-plane and out-of-plane neighbors and 
 (ii) the Hubbard-U correction. To take into account the first effect 
 we note that the $d_{xy}$ wavefunctions only interact strongly with in-plane 
 neighbors which are all functions centered around Ru$^{4+}$ ions. 
 In contrast, $d_{xz,yz}$ wavefunctions interact in-plane with 
 centers around Ru$^{4+}$ ions while out-of-plane they interact with 
 functions centered on Ti$^{4+}$ ions
 mediated by O atoms [see Fig.~\ref{fig:model}(e)]. 
 Since Ru$^{4+}$ ions are more electronegative than Ti$^{4+}$ ones,  
 the apical oxygens are polarized towards the Ru$^{4+}$.
 This translates into an increase of covalency and, therefore,
 a decrease in the energy of the bonding states and an increase of the
 antibonding levels, denoted as $\Delta E_{\rm{cov}}$ 
 in Fig. ~\ref{fig:model}(c).
 Finally, we include the effect of the $U_{\rm eff}$ value to take into 
 account magnetism. LDA+U theory predicts that orbital energies are 
 shifted by $U_{\rm eff}(1/2-\lambda)$ where $\lambda$ is the occupation 
 of the orbital. 
 Using this formula, and assuming that the majority spin 
 $t_{2g}$ levels are full, while the minority spin 
 $d_{xz}$, $d_{yz}$ and $d_{xy}$ contain respectively 0.5, 0.5 and 
 0.0 electrons, we 
 find the shifts depicted in Fig.~\ref{fig:model}(d). Qualitatively, 
 we find the bands to be in good agreement with those obtained in the full 
 first-principles calculations [see Fig.~\ref{fig:dos-sro}(b)]. 

 In summary, our first principles simulations predict the appearance of a
 two-dimensional, half-metallic, ferromagnetically ordered 
 electron gas at the insulator/ultrathin-metal-film interfaces
 in (STO)$_{5}$/(SRO)$_{1}$ superlattices. 
 The 2DEG is extremely confined at Ru 4$d$ orbitals.
 At odds with previous realizations of 2DEG, where the main mechanism
 behind localization is of electrostatic nature~\cite{Stengel-11.2,Delugas-11},
 we propose the larger electronegativity of Ru to be the main cause
 for the extreme confinement.
 The half-metallic properties and the concomitant FM is an intrinsic
 property due to enhanced electron correlation, and not to extrinsic 
 properties (O vacancies) as postulated in LAO/STO 2DEG~\cite{Brinkman-07}.
 Our results encourage experimental verification, 
 to check the usefulness of these superlattices in 
 magnetoresistance or spintronic devices.

 Financial support from grants
 FIS2009-12721-C04-02, and
 CP-FP 228989-2 OxIDes.
 Calculations were performed on the computers at the ATC group
 and on the Altamira Supercomputer of the RES.


%

\end{document}